\documentclass[epj]{svjour}

\usepackage{rotating}
\usepackage{graphicx}

\newcommand{\la}{\langle}
\newcommand{\ra}{\rangle}

\begin{document}

\title{Many-body Green's function theory for the magnetic
reorientation of thin ferromagnetic films}
\author{P. Fr\"obrich\inst{1,2}, P.J. Jensen\inst{1,2},
P.J. Kuntz\inst{1},  and A. Ecker\inst{1}}
\institute{Hahn-Meitner-Institut Berlin, Glienicker Stra{\ss}e 100,
D-14109 Berlin, Germany, \and  also: Institut f\"ur Theoretische
Physik, Freie Universit\"at Berlin\\ Arnimallee 14, D-14195 Berlin,
Germany } 
\date{Received: } 
\titlerunning{Many-body Green's
function theory for the magnetic reorientation \ldots}
\authorrunning{P. Fr\"obrich \em et al.\em}  

\abstract{  The field-induced reorientation of the magnetization of
ferromagnetic films is treated within the framework of many-body
Green's function theory by considering all components of the
magnetization.  We present a new method for the calculation of
expectation values in terms of the eigenvalues and eigenvectors of the
equations of motion matrix for the set of Green's functions. This
formulation allows a straightforward extension of the monolayer case
to thin films with many layers and for arbitrary spin and moreover
provides a practicable  procedure for numerical computation. The model
Hamiltonian includes a Heisenberg term, an external magnetic field, a
second-order uniaxial single-ion anisotropy, and the magnetic
dipole-dipole coupling. We utilize the Tyablikov (RPA) decoupling for
the exchange interaction terms and the Anderson-Callen decoupling for
the anisotropy terms.  The dipole coupling is treated in the
mean-field approximation, a procedure which we demonstrate to be a
sufficiently good approximation for realistic coupling strengths.  We
apply the new method to monolayers with spin $S\geq 1$ and to
multilayer systems with $S=1$.  We compare some of our results to
those where mean-field theory (MFT) is applied to all interactions,
pointing out some significant differences.}  \PACS{
{75.10.Jm}{Quantized spin models} \and  {75.30.Ds}{Spin waves} \and
{75.70.Ak}{Magnetic properties of monolayers and thin films}}
\maketitle

\section{Introduction}
In this paper, we extend our earlier investigations
\cite{EFJK99,FJK00} on the reorientation of the magnetization of a
ferromagnetic monolayer with spin $S=1$ to multilayer systems with
arbitrary spin, $S$.  The components of the magnetization as functions
of temperature and film thickness are calculated within the framework
of a many-body Green's function theory, allowing the {\em direct}
calculation of the magnetic orientation. Furthermore, we derive and
apply a non-perturbative expression for the temperature dependence of
the second-order single-ion anisotropy by minimizing the free energy
with respect to this orientation angle.

For convenience, mean-field theory (MFT) is often applied to such
problems, either by diagonalisation of a single-particle Hamiltonian
\cite{JeB98}, or by a thermodynamic perturbation theory \cite{theo}.
We emphasize that this approximation completely neglects collective
excitations (spin waves) \break  which are known to have a much
greater influence on the magnetic properties of 2D systems than on 3D
bulk properties.  In fact, recent calculations on a trilayer system
\cite{Je99} demonstrate that MFT is incapable of accounting for the
induced magnetization observed in coupled layers unless an
unrealistically large interlayer coupling is postulated.  On the other
hand, the many-body Green's function techniques, which take the
collective excitations  approximately into account, can explain the
experimental observations assuming an interlayer exchange coupling of
reasonable size. We point out that the dependence of the magnetization
and the Curie temperature on the film thickness has also been studied
within Green's function theory \cite{SN99}.  Also, in references
\cite{Bru91} and \cite{SW92} Green's function techniques are applied
to magnetization problems. However, in all these references only a
single component of the magnetization is considered. Therefore, a
reorientation of the magnetization as a function of temperature, film
thickness, or magnetic field cannot be calculated.  A reorientation of
the magnetization is considered in reference \cite{EM91}, where
Green's functions are used after a Holstein-Primakoff mapping of the
spin operators to bosons. In order to solve the non-Hermitian
eigenvalue problem, right and left eigenvectors are applied for its
solution, similar to what is done in the present paper. The theory is,
however, only valid at low temperatures.  Another method for the
treatment of the magnetic reorientation for all temperatures is a
Schwinger-Boson theory \cite{Timm00}, as an alternative to the Green's
function method of this paper.

In the present work, we treat the field-induced reorientation of the
magnetization for all temperatures of interest. Since expectation
values of all three components of the spin operator are considered, a
corresponding set of Green's functions must be defined. We introduce a
new method for the calculation of the expectation values, which
utilizes not only the eigenvalues but also the eigenvectors of the
(non-symmetric) matrix governing the equations of motion for the
Green's functions.  This formulation is more compact than the usual
one \cite{FJK00} and, most importantly, suggests a practicable way of
treating the multilayer case for arbitrary spin. We make no attempt to
go beyond the Tyablikov (Random Phase Approximation: RPA) decoupling
for the exchange terms, since we have shown in reference \cite{EFJK99}
that this decoupling scheme for a monolayer with spin $S=1/2$ compares
well indeed with an `exact' Quantum Monte Carlo (QMC) calculation.
The single-ion anisotropy term is decoupled with the Anderson-Callen
method \cite{AC64}, because, as shown in \cite{FJK00}, other
single-ion decoupling schemes, e.g.\ that of Lines \cite{Lin67}, lead
to difficulties when calculating the magnetic reorientation.
Furthermore, we include the magnetic dipole coupling, which is treated
within a simplified (non-dispersive) approximation, which corresponds
to its mean field treatment. The dipole coupling was not considered in
our earlier work \cite{FJK00}, which was restricted to the case of a
monolayer with spin $S=1$.

The paper is organized as follows. In Section 2 the Green's function
formalism is outlined. The eigenvector method is then presented for
the monolayer case and it is shown that this leads to a transparent
extension to the case of many layers.  Section 3 deals with the
results. First the effect of the dipole coupling on the magnetic
reorientation is discussed using the monolayer with spin $S=1$ as an
example. Secondly, we discuss  monolayers with spins $S>1$. Thirdly
the formalism is applied to the case of many layers.  Finally Section
4 contains a discussion of the results and an outlook for further
investigations.  In Appendix A, different approximations for the
magnetic dipole coupling are investigated.  Details of the formalism
for $S>1$ are derived in Appendix B.

\section{The Green's function formalism}
In order to study the field-induced magnetic reorientation of a
ferromagnetic thin film we investigate a spin Hamiltonian consisting
of an isotropic Heisenberg exchange interaction, $J_{kl}$, between
nearest neighbour lattice sites, a second-order single-ion lattice
anisotropy with strength, $K_{2,k}$, the magnetic dipole coupling with
strength, $g_{kl}$, and an external magnetic field,
${\bf{B}}=(B^x,B^y,B^z)$,
\begin{eqnarray}
{\cal H}&=&-\frac{1}{2}\sum_{<kl>}J_{kl}\;{\bf S}_k\cdot{\bf S}_l
-\sum_kK_{2,k}(S_k^z)^2 \nonumber\\ 
&& -\sum_k\Big(\frac{1}{2}B^-S_k^++\frac{1}{2}B^+S_k^-
+B^zS_k^z\Big) \\ 
& & +\frac{1}{2}\sum_{kl}\frac{g_{kl}}{r_{kl}^5}\Big(r_{kl}^2\, 
{\bf S}_k\cdot{\bf S}_l-3({\bf S}_k\cdot{\bf r}_{kl})({\bf S}_l
\cdot{\bf r}_{kl})\Big)\;. \nonumber \label{1} \end{eqnarray}

Here the notation $S_i^\pm=S_i^x\pm iS_i^y$ and $B^\pm=B^x\pm iB^y$ is
introduced, $k$ and $l$ are lattice site indices, and  $\la kl\ra$
indicates summation over nearest neighbours only.  Each layer is
assumed to be ferromagnetically ordered (collinear magnetization),
whereas the magnetization of different layers need not to be
collinearly aligned. Furthermore, inhomogeneous systems can be
considered which are characterized by different layer-dependent
coupling constants and magnetic moments.  We do not include a fourth
order uniaxial anisotropy term, $-\sum_k K_{k,4}(S_k^z)^4$, because it
is difficult to find a proper decoupling of this term in the equations
of motion for the Green's functions. This means that the formalism of
the present paper is only adequate for the physical situation in which
such a term is of no importance.

As in reference \cite{FJK00}, we introduce the set of thermal Green's
functions in the spectral representation
\begin{equation}
G_{ij(\eta)}^{\alpha,mn}(\omega)=\la\la
S_i^\alpha;(S_j^z)^m(S_j^-)^n\ra\ra_\omega\;; \ \ \alpha=+,-,z \ \ ,
\label{2} \end{equation}
where $\omega$ denotes the energy, and $\eta=\pm1$ refers to the
commutator ($\eta=-1$) or anti-commutator ($\eta=+1$) Green's
functions, respectively; $n\ge1$ and $m\ge0$ are integers, $i$ and $j$
denote lattice sites.  In order to obtain a closed set of equations of
motion for the Green's functions, we treat the exchange term by a
generalized Tyablikov (RPA) \cite{Tya67} decoupling, and the
anisotropy term by the Anderson-Callen decoupling \cite{AC64}.

In our previous work the corresponding thermal correlation functions
\begin{equation}
C_{ij}^{mn,\alpha}=\la(S^z_j)^m(S^-_j)^nS_i^{\alpha}\ra
\label{2a} \end{equation}
have been obtained by applying the spectral theorem \cite{Tya67}.
Because of vanishing eigenvalues it is important to use the spectral theorem
including the term obtained from the anti-commutator Green's functions
\begin{equation}
D^{mn}_{ij}=\frac{1}{2}\lim_{\omega \to 0}G_{ij(\eta=+1)}^{mn}\,.
\label{2b} \end{equation}
Together with the so-called regularity conditions, which are derived
from the fact that the spectral representation of the commutator
Green's function must be regular for $\omega=0$, we have derived a set
of coupled equations for the correlation functions. The solution
yields the components of the magnetization, thus determining 
{\it directly} the reorientation angle of the magnetization induced by the
applied external field.  In the present paper, we rederive these
equations by a method which utilizes the eigenvectors as well as the
eigenvalues of the matrix determining the Green's functions.  This
more compact formulation furnishes a practicable way to treat the
multilayer case and general spin quantum numbers $S$. It is
didactically advantageous to demonstrate this new method first for a
monolayer; the generalization to the multilayer case then follows in a
straightforward and transparent way.

\subsection{The eigenvector method for the monolayer}

The equations of motion for the Green's functions in the spectral
representation read
\begin{equation}
\omega\;G_{ij(\eta)}^{\alpha,mn}(\omega)=A_{ij(\eta)}^{\alpha,mn}+\la\la
[S_i^\alpha,{\cal H}]_{-1};(S_j^z)^m(S_j^-)^n\ra\ra \,,
\label{3} \end{equation}
with the inhomogeneities
\begin{eqnarray}
A_{ij(\eta)}^{\alpha,mn}&=&\la[S_i^\alpha,(S_j^z)^m(S_j^-)^n]_{\eta}\ra 
\nonumber \\
&=&\la S_i^\alpha(S_j^z)^m(S_j^-)^n+\eta(S_j^z)^m(S_j^-)^nS_i^\alpha\ra\,,
\label{4} \end{eqnarray}
where $\la\cdots\ra={\rm Tr}(\cdots e^{-\beta\cal{H}})$ with 
$\beta=1/k_BT$ and $k_B$ Boltzmann's constant, and $\eta=+1$ or $-1$,
respectively.

The higher Green's functions in equation (\ref{3}) due to the exchange
interaction term are decoupled by a generalized Tyablikov (RPA)
decoupling \cite{Tya59}
\begin{equation}
\la\la S_i^{\alpha}S_k^{\beta};(S_j^z)^m(S_j^-)^n\ra\ra\simeq
\la S_i^{\alpha}\ra\;G_{kj}^{\beta,mn}+\la S_k^{\beta}\ra\;
G_{ij}^{\alpha,mn}\;. \label{4a} \end{equation}
In reference \cite{FJK00} the proper inclusion of the single-ion
aniso\-tropy with Green's function techniques was thoroughly discussed
in connection with the magnetic reorientation. Accordingly, we choose
the Anderson-Callen \cite{AC64,FJK00} decoupling for the treatment of
the anisotropy terms:
\begin{eqnarray}
&& \la\la S_i^{\pm}S_i^z+S_i^zS_i^{\pm}\ra\ra \nonumber \\ 
&& \simeq 2\la S_i^z\ra \Big( 1-\frac{1}{2S^2}[S(S+1) 
-\la S_i^zS_i^z\ra] \Big)G_{ij}^{\pm,mn}\,. \label{4b}
\end{eqnarray}

Because we are interested in laterally periodic systems we perform a
Fourier transformation to the two-dimensional wave vector space 
${\bf k}$.  Introducing vectors for the Green's functions, 
${\bf G}_{\eta}^{mn}({\bf k},\omega)$, and for the inhomogeneities, 
${\bf A}_{\eta}^{mn}$,
\begin{equation}
{\bf G}_{\eta}^{mn}({\bf{k},\omega})=\left( \begin{array}{c}
G_{\eta}^{+,mn}({\bf{k}},\omega) \\ G_{\eta}^{-,mn}({\bf{k}},\omega)  \\
G_{\eta}^{z,mn}({\bf{k}},\omega) \end{array} \right)\,, \hspace{0.3cm}
{\bf A}_{\eta}^{mn}= \left( \begin{array}{c} A_\eta^{+,mn} \\ 
A_\eta^{-,mn} \\ A_\eta^{z,mn} \end{array} \right) \,, \label{5} 
\end{equation}
the equations of motion, which are derived in detail in reference
\cite{FJK00}, can be written in a compact form
\begin{equation}
(\omega\,{\bf 1}-{\bf \Gamma})\;{\bf G}_{\eta}^{mn}={\bf A}_{\eta}^{mn},
\label{6} \end{equation}
where {\bf 1} is the unit matrix and the {\it non-symmetric} matrix 
${\bf\Gamma}$ is given by
\begin{equation}	
{\bf \Gamma} = \left( \begin{array}
{@{\hspace*{3mm}}c@{\hspace*{5mm}}c@{\hspace*{5mm}}c@{\hspace*{3mm}}}
\;\;\;\;\tilde{H^z} & 0 & -H^+ \\ 0 & -\tilde{H^z}& \;\;\;H^- \\
-\frac{1}{2}H^- & \;\frac{1}{2}H^+ & 0
\end{array} \right) \;,\label{7} \end{equation}
with the abbreviations
\begin{eqnarray}
H^\alpha&=&B^\alpha+\la S^\alpha\ra\,J\,(q-\gamma_{\bf k})\,,
\qquad \alpha=+,-,z\,, \nonumber \\
\tilde{H}^z&=&H^z+K_2\,\Phi\;=\;Z+ \la S^z\ra\,J\,(q-\gamma_{\bf k})\,,
\nonumber\\ Z&=&B^z+K_2\,\Phi\,, \label{8} \\
\Phi&=&2\la S^z\ra \Big(1-\frac{1}{2S^2}[S(S+1)-
\la S^zS^z\ra]\Big)\;. \nonumber \end{eqnarray}
For a square lattice with a lattice constant taken to be unity, one
obtains $\gamma_{\bf k}=2(\cos k_x+\cos k_y)$, and $q=4$ is the number
of nearest neighbours.  Note that ${\bf A}_\eta^{mn}={\bf
A}_\eta^{mn}({\bf k})$ depends on the wave vector ${\bf k}$ for
$\eta=1$ but not for $\eta=-1$.

Similar to the exchange coupling, the long-range dipole coupling also
induces a momentum dependence into the magnon dispersion relation
$\omega({\bf k})$. Due to the oscillating lattice sums the
consideration of the {\bf k}- dependence of this coupling obtained
e.g.\ in RPA is fairly complicated and  time consuming.  Therefore we
accept for the present calculations an approximate description of the
dipole coupling in the dispersion relation, in particular its {\bf k}-
dependent terms are neglected, which is equivalent to its mean field
approximation. In Appendix A, we show that for dipole interactions
small compared to the exchange coupling, which is the case for the
ferromagnetic $3d$- transition metals, this approximation is
satisfactory.  The approximation of Appendix A merely leads to a
renormalization of the external magnetic field components $B^\pm$ and
$B^z$, which for the $i$th atomic layer in the case of a multilayer
with $N$ layers reads
\begin{eqnarray}
B^{\pm}_i & \to & B^{\pm}+\sum_{j=1}^N\;g_{ij}\;\la S_j^{\pm}\ra\;
T^{|i-j|} \;, \nonumber\\
B^z_i & \to & B^z-2\sum_{j=1}^N\;g_{ij}\;\la S_j^z\ra\;
T^{|i-j|} \;, \label{39} \end{eqnarray}
where the lattice sums for a two-dimensional square lattice are given by
($n=|i-j|$)
\begin{equation}
T^n = \sum_{lm}\frac{l^2-n^2}{(l^2+m^2+n^2)^{5/2}} \;. \label{25}
\end{equation}
The indices ($lm$)  run over all sites of the square $j$-th layer,
excluding the terms with $l^2+m^2+n^2=0$.  For the monolayer ($N=1$) one
has $i=j$, and one obtains in particular $T^0\simeq 4.5165$.  As can
be seen from equations (\ref{39}), the dipole coupling reduces the
effect of the external field component in $z$-direction and enhances
the effect of the transversal field components $B^\pm$.

We now introduce a transformation which diagonalizes the
matrix ${\bf \Gamma}$
\begin{equation}
{\bf L\;\Gamma\;R}={\bf\Omega}= \left( \begin{array}
{@{\hspace*{3mm}}c@{\hspace*{5mm}}c@{\hspace*{5mm}}c@{\hspace*{3mm}}}
\omega_0 & 0 & 0 \\ 0 & \omega_+ & 0 \\ 0 & 0 & \omega_-
\end{array} \right) \;, \label{9} \end{equation}
where the eigenvalues turn out to be $\omega_0=0,\ \omega_\pm=\pm
E_{\bf k}$ with $E_{\bf k}^2=H^+H^-+\tilde{H^z}\tilde{H^z}$.  The
transformation matrix ${\bf R}$ and its inverse ${\bf R}^{-1}={\bf L}$
are obtained from the right eigenvectors of ${\bf\Gamma}$ as columns
and from the left eigenvectors as rows, respectively.  These matrices
are normalised to unity: $\bf {LR=RL=1}$.  Note that due to the
non-symmetric matrix ${\bf\Gamma}$, equation (\ref{7}), one has in
general ${\bf R}^{-1}\neq {\bf R^T}$, ${\bf R^T}$ being the transposed
matrix.

For the monolayer, the transformation matrices can be constructed
analytically; the right eigenvectors are arranged so that the columns
$1$, $2$, and $3$ correspond to the eigenvalues $0$, $+E_{\bf k}$, and
$-E_{\bf k}$, respectively:
\begin{eqnarray}
{\bf R} & = & \left( \begin{array}
{@{\hspace*{1mm}}c@{\hspace*{3mm}}c@{\hspace*{3mm}}c@{\hspace*{1mm}}}
{H^+}/{\tilde{H^z}} & -({\tilde{H^z}+E_{\bf k}})/{H^-} &
-({\tilde{H^z}-E_{\bf k}})/{H^-} \\
{H^-}/{\tilde{H^z}} & -({\tilde{H^z}-E_{\bf k}})/{H^-} &
-({\tilde{H^z}+E_{\bf k}})/{H^-} \\
1 & 1 & 1  \end{array} \right)\;. \nonumber \\ \label{10}
\end{eqnarray}
Similarly, the left eigenvectors are arranged so that rows $1$, $2$,
and $3$ correspond to the eigenvalues $0$, $+E_{\bf k}$, and 
$-E_{\bf k}$, see equation (\ref{11}) above. 

\begin{figure*}[t]
\begin{equation}
{\bf L }=\frac{1}{4E_{\bf k}^2}
\left( \begin{array}
{@{\hspace*{3mm}}c@{\hspace*{5mm}}c@{\hspace*{5mm}}c@{\hspace*{3mm}}}
2H^-\tilde{H^z} & 2H^+\tilde{H^z} & 4\tilde{H^z}\tilde{H^z} \\
-H^-(E_{\bf k}+\tilde{H^z}) & \;\;\;H^+(E_{\bf k}-\tilde{H^z}) & 2H^-H^+ \\
\;\;\;H^-(E_{\bf k}-\tilde{H^z}) & -H^+(E_{\bf k}+\tilde{H^z}) &
2H^-H^+\ \end{array} \right)\;. \label{11} \end{equation}
\hrulefill \\ \vspace{-0.5cm}  \end{figure*}
Multiplying the equation of motion (\ref{6}) from the left by ${\bf L}$ and
inserting ${\bf 1}={\bf RL}$ one finds
\begin{equation}
(\omega\;{\bf 1}-{\bf \Omega})\,{\bf L}\,{\bf G}_{\eta}^{mn}={\bf L}\,
{\bf A}_{\eta}^{mn}\;. \label{12} \end{equation}
Defining ${\cal G}_{\eta}^{mn}\equiv{\bf L}\,{\bf G}_{\eta}^{mn}$ and
${\cal A}_{\eta}^{mn}\equiv{\bf L}\,{\bf A}_{\eta}^{mn}$
one obtains 
\begin{equation}
(\omega\;{\bf 1}-{\bf \Omega})\;{\cal G}_{\eta}^{mn}=
{\bf {\cal A}}_{\eta}^{mn}. \label{13} \end{equation}
${\cal G}_{\eta}^{mn}$ is a new vector of Green's functions, each component
$\tau$ of which has only a single pole
\begin{equation}
{\cal G}_{\eta}^{mn,\tau}=\frac{{\cal A}_{\eta}^{mn,\tau}}{\omega -
\omega_{\tau}}\,. \label{14} \end{equation}
This allows us to apply the spectral theorem to each component
separately.  We introduce the vectors ${\cal C}^{mn}\equiv{\bf
L}\,{\bf C}^{mn}$ for the correlations, and ${\cal D}^{mn}\equiv{\bf
L}{\bf D}^{mn}$ for the correction to the spectral theorem in case of
a vanishing eigenvalue. Application of the spectral theorem
\cite{Tya67} to the $\tau$th component of the single-pole Green's
function of equation (\ref{13}) then yields
\begin{equation}
{\cal
C}^{mn,\tau}=\frac{{\cal A}_{\eta}^{mn,\tau}}{e^{\beta\omega_{\tau}}+\eta}
+\frac{1}{2}(1-\eta)\;{\cal D}^{mn,\tau}\;, \label{15} \end{equation}
where
\begin{eqnarray}
{\cal D}^{mn,\tau}&&=\frac{1}{2}\;\lim_{\omega\rightarrow 0}\;\omega\;
{\cal G}_{\eta=+1}^{mn,\tau} \nonumber \\
& & =\frac{1}{2}\lim_{\omega\rightarrow 0}
\;\frac{\omega\;{\cal A}_{\eta=+1}^{mn,\tau}}{\omega-\omega_{\tau}}
=\frac{1}{2}\;\delta_{\tau 0}\;{\cal A}_{\eta=+1}^{mn,\tau}\,.
\label{16} \end{eqnarray}
Here $\delta_{\tau 0}$ is the Kronecker symbol which ensures that
${\cal D}^{mn,\tau}$ has a non-zero value only
if $\tau$ refers to the component with eigenvalue zero.

Denoting ${\bf L}^0$ as the left eigenvector corresponding to eigenvalue zero,
we find
\begin{eqnarray}
&&{\cal D}^{mn,0}=\frac{1}{2}\;{\cal A}_{\eta=+1}^{mn,0}=\frac{1}{2}\;
{\bf L}^0\;{\bf A}_{\eta=+1}^{mn} \\ 
&& =\frac{1}{2}\;{\bf L}^0\;({\bf A}_{\eta=-1}^{mn}+2\,{\bf C}^{mn})= 
\sum_{\alpha}L^0_{\alpha}\;C^{mn,\alpha}={\cal C}^{mn,0}\,. \nonumber 
\label{17} \end{eqnarray}
Here we have used the relation between the commutator and anticommutator
inhomogeneities, ${\bf A}_{+1}^{mn}({\bf k})={\bf A}_{-1}^{mn}+2{\bf C}_{\bf
k}^{mn}$, and the regularity
condition for the commutator Green's function for $\omega\rightarrow 0$
\begin{equation}
{\cal A}_{\eta=-1}^{mn,0}=\sum_\alpha L_\alpha^0\;A_{\eta=-1}^{\alpha,mn}
=0\,. \label{18} \end{equation}
One sees explicitly, when inserting the left eigenvector of equation (\ref{11})
belonging to eigenvalue zero, that
\begin{eqnarray}
\sum_\alpha L_{\alpha}^0\;A_{\eta=-1}^{\alpha,mn}&&= 
\frac{1}{2E_{\bf k}^2} \times \\ 
\;\bigg(H^-\tilde{H^z}&&,H^+\tilde{H^z},2\tilde{H^z}\tilde{H^z}\bigg)
 \left( \begin{array}{ccc} A_{\eta=-1}^{+,mn} \\ A_{\eta=-1}^{-,mn} \\
A_{\eta=-1}^{z,mn} \end{array} \right)=0 
\nonumber \label{19} \end{eqnarray}
are the regularity conditions of equation (17) of reference \cite{FJK00}, see
also Appendix B of this reference.

The components of the correlation vector
${\cal C}^{mn}$ for $\omega_{\tau}\neq 0$ are of the form
\begin{equation}
{\cal C}^{mn,\tau}=\frac{{\cal
A}^{mn,\tau}_{\eta=-1}}{e^{\beta\omega_{\tau}}-1}+{\cal D}^{mn,0}\;.
\label{20} \end{equation}
The original correlation vector ${\bf C}^{mn}$ can be recovered
from ${\cal C}^{mn}$ by multiplying from the left with ${\bf R}$; one
obtains a compact expression by first defining a matrix ${\cal L}$ in
terms of row-vectors ${\cal L}^{\tau}$ corresponding to the
row-vectors of ${\bf L}\;$:
\begin{eqnarray}
{\cal L}^{\tau} & = & \frac{1}{e^{\beta\omega_\tau}-1}\;{\bf L}^{\tau}\;,
\hspace{0.5cm} (\tau\neq 0) \\
{\cal L}^0 & = & {\bf 0} \;, \label{21}
\end{eqnarray}
so that
\begin{equation}
{\bf C}^{mn}={\bf R}\;{\cal C}^{mn}={\bf R}\;{\cal L}\;{\bf A}_{\eta=-1}^{mn}
+{\bf R}\;{\cal D}^{mn}\;. \label{22} \end{equation}
The final equation determining the correlation vector is
\begin{equation}
{\bf C}^{mn}={\bf R}\;{\cal L}\;{\bf A}_{\eta=-1}^{mn}+
{\bf R}^0\;{\bf L}^0\;{\bf C}^{mn}\;. \label{23}
\end{equation}
The product ${\bf R}^0\,{\bf L}^0$ is a projection operator onto the
subspace belonging to the eigenvector corresponding to
$\omega_\tau=0$, so that the term ${\bf R}^0\,{\bf L}^0\,{\bf C}^{mn}$
is the projection of the correlation vector onto this subspace.  This
interpretation carries over to the $N-$layer case, where, it will be
seen, there is an $N-$dimensional space corresponding to the zero
eigenvalues. It is important to stress that this equation is not
complete but must be supported by the regularity conditions
(\ref{19}).  Inserting the matrices ${\bf R}$ and ${\bf L}$ from
equations (\ref{10}) and (\ref{11}) one sees that the $z$-component of
this equation is exactly equation (27) of reference \cite{FJK00}. One
could equivalently use the ($+$) or ($-$)-components of this equation,
which can be proved to give the same results.

In reference \cite{FJK00} we have investigated only spin $S=1$. In
this case, it is sufficient to use the equations for $(mn)=(01),(02)$
and $(11)$.  For general spin $S$, all regularity conditions with
$(m+n)\leq 2S$ have to be taken into account. They form a set of
linear equations which allow one to express all correlations ocurring
in equation (\ref{23}) in terms of the moments $\la(S^z)^p\ra$
($p=1,\ldots,2S+1$). This leads to $2S+1$ equations for the moments
$\la(S^z)^{p}\ra$, which can be reduced to $2S$ equations by
expressing the highest moment in terms of lower ones using the
condition $\prod_{M_S=-S}^S(S^z-M_S)$=0. Note that only the first two
equations have to be iterated because in the dispersion relation only
$\la S^z\ra$ and $\la S^zS^z\ra$ occur. For more details, see Appendix
B.
\begin{figure*}[t] 
\setcounter{equation}{40}
\begin{eqnarray}
{\bf L}^0_i\;{\bf \Gamma}_{ij}&=&\frac{\tilde{H_i^z}}{2E_{\bf k}^2} \;
J_{ij}(H_i^-,H_i^+,2\tilde{H_i^z}) \left( \begin{array}
{@{\hspace*{3mm}}c@{\hspace*{5mm}}c@{\hspace*{5mm}}c@{\hspace*{3mm}}}
-\la S_i^z\ra & 0 & \;\;\;\la S_i^+\ra \\
0 & \;\;\la S_i^z\ra & -\la S_i^-\ra \\
\frac{1}{2}\la S_i^-\ra & -\frac{1}{2}\la S_i^+\ra & 0
\end{array} \right) \nonumber \\
&=&\frac{\tilde{H_i^z}}{2E_{\bf}^2}\;J_{ij}
\Big(-H_i^-\la S_i^z\ra +\tilde{H_i^z}\la S_i^-\ra,
\;H_i^+\la S_i^z\ra -\tilde{H_i^z}\la S_i^+\ra,
\;H_i^-\la S_i^+\ra -H_i^+\la S_i^-\ra \Big) 
= (0,0,0)\;. \label{36} \end{eqnarray}
\hrulefill \\ \vspace{-0.5cm} 
\end{figure*}

\subsection{Multilayers}
\setcounter{equation}{30}
Having established the formalism for the monolayer, it is now
relatively easy to generalize to the multilayer case.  For a
ferromagnetic film with $N$ layers the $3N$ equations of motion for
the $3N$ dimensional Green's function vector ${\bf G}^{mn}$ read
\begin{equation}
(\omega\,{\bf 1}-{\bf \Gamma})\,{\bf G}^{mn}={\bf A}^{mn}\;,
\label{26} \end{equation}
where ${\bf 1}$ is the $3N \times 3N$ unit matrix, and the Green's
function and inhomogeneity vectors consist of $N$ three-dimensional
subvectors which are characterized by the layer indices $i$ and $j$
\begin{equation}
{\bf G}_{ij}^{mn}({\bf{k},\omega})\  =
\left( \begin{array}{c}
G_{ij}^{+,mn}({\bf{k}},\omega) \\ G_{ij}^{-,mn}({\bf{k}},\omega)  \\
G_{ij}^{z,mn}({\bf{k}},\omega)
\end{array} \right), \hspace{0.5cm}
{\bf A}_{ij}^{mn} {=}
 \left( \begin{array}{c} A_{ij}^{+,mn} \\ A_{ij}^{-,mn} \\
A_{ij}^{z,mn} \end{array} \right) \;. \label{27} \end{equation}

The equations of motion are then expressed in terms of these layer
vectors, and $3\times 3 $ submatrices ${\bf \Gamma}_{ij}$ of the
$3N\times 3N$ matrix ${\bf\Gamma}$ 
\begin{equation}
\left[ \omega {\bf 1}-\left( \begin{array}{cccc}
{\bf\Gamma}_{11} & {\bf\Gamma}_{12} & \ldots & {\bf\Gamma}_{1N} \\
{\bf\Gamma}_{21} & {\bf\Gamma}_{22} & \ldots & {\bf\Gamma}_{2N} \\
\ldots & \ldots & \ldots & \ldots \\
{\bf\Gamma}_{N1} & {\bf\Gamma}_{N2} & \ldots & {\bf\Gamma}_{NN}
\end{array}\right)\right]\left[ \begin{array}{c}
{\bf G}_{1j} \\ {\bf G}_{2j} \\ \ldots \\ {\bf G}_{Nj} \end{array}
\right]=\left[ \begin{array}{c}
{\bf A}_{1j}\delta_{1j} \\ {\bf A}_{2j}\delta_{1j} \\ \ldots \\
{\bf A}_{Nj}\delta_{1j} \end{array}
\right] \;, \label{28} \end{equation}
$j=1,\ldots,N$. After applying the decoupling procedures  (\ref{4a}) and
(\ref{4b}), the  ${\bf\Gamma}$ matrix reduces to a band matrix with
zeros in the ${\bf\Gamma}_{ij}$ sub-matrices, when $j>i+1$ and
$j<i-1$.  The diagonal sub-matrices ${\bf\Gamma}_{ii}$ are of size
$3\times 3$ and turn out to have the same structure as the matrix
which characterizes the monolayer, see equation (\ref{7}):
\begin{equation}
 {\bf \Gamma}_{ii}= \left( \begin{array}
{@{\hspace*{3mm}}c@{\hspace*{5mm}}c@{\hspace*{5mm}}c@{\hspace*{3mm}}}
\;\;\;\tilde{H^z_i} & 0 & -H^+_i \\ 0 & -\tilde{H^z_i} & \;\;\;H^-_i \\
-\frac{1}{2}H^-_i & \;\frac{1}{2}H^+_i & 0 \end{array} \right) \ .
\label{29} \end{equation}
In particular one of the eigenvalues of ${\bf\Gamma}_{ii}$ vanishes.
 The matrix elements of ${\bf\Gamma}_{ii}$ contain additional terms
 due to the exchange interaction between the atomic layers, the dipole
 coupling is contained in the field components $B^\alpha_i$, see 
 equation (\ref{39}),
\begin{eqnarray}
H^\alpha_i&=&B^\alpha_i+\la S_i^\alpha\ra\,J_{ii}\,(q-\gamma_{\bf k})
+J_{i,i+1}\la\,S_{i+1}^{\alpha}\ra \nonumber \\
&& +J_{i,i-1}\,\la S_{i-1}^{\alpha}\ra \,, \nonumber \\
\tilde{H}^z_i&=&H^z_i+K_{2,i}\;\Phi_i\;=\;Z_i+ \la S_i^z\ra\,
J_{ii}\,(q-\gamma_{\bf k}), \nonumber \\
Z_i&=&B^z_i+J_{i,i+1}\,\la S_{i+1}^{z}\ra
+J_{i,i-1}\,\la S_{i-1}^{z}\ra+K_{2,i}\;\Phi_i \,, \nonumber \\
\Phi_i&=&2\la S_i^z\ra\Big(1-\frac{1}{2S^2}[S(S+1)-\la S_i^zS_i^z\ra]\Big) 
\;, \label{30} \end{eqnarray}
and $\alpha=+,-,z$. The $3\times 3$ non-diagonal sub-matrices 
${\bf\Gamma}_{ij}$ for $j=i\pm 1$ are of the form
\begin{equation}
 {\bf \Gamma}_{ij} = \left( \begin{array}
{@{\hspace*{1mm}}c@{\hspace*{3mm}}c@{\hspace*{3mm}}c@{\hspace*{1mm}}}
-J_{ij}\la S_i^z\ra & 0 & \;\;\;J_{ij}\la S_i^+\ra \\
0 & \;\;J_{ij}\la S_i^z\ra & -J_{ij}\la S_i^-\ra \\
\frac{1}{2}J_{ij}\la S_i^-\ra &
-\frac{1}{2}J_{ij}\la S_i^+\ra & 0 \end{array} \right) \;.
\label{31} \end{equation}
We now demonstrate that there is a left eigenvector of ${\bf\Gamma}$
corresponding to eigenvalue zero with the structure
\begin{equation}
{\bf L}^0=(0,\ldots,0,{\bf L}^0_i,0,\ldots,0) \;, \label{32} \end{equation}
where
\begin{equation}
{\bf L}_i^0=(L_{i1}^0,L_{i2}^0,L_{i3}^0)=\frac{1}{2E_{\bf k}^2} \;
\Big(H_i^-\tilde{H^z_i},\;H_i^+\tilde{H^z_i},\;2\tilde{H^z_i}\tilde{H^z_i}
\Big) \;. \label{33} \end{equation} 
This is immediately clear for the diagonal elements, because they have
the same structure as the monolayer matrix, equation (\ref{7}).  To
prove this also for the non-diagonal matrix elements one needs the
regularity condition (\ref{19}) for layer $i$:
\begin{equation}
\sum_{\alpha}\,L_{i\alpha}^0\;A_i^{\alpha,mn}=0\,, \label{34} \end{equation}
for $m=0$, $n=1$.
With $A_i^{+,01}=2\la S_i^z\ra$, $A_i^{-,01}=0$, and
$A_i^{z,01}=-\la S_i^-\ra$ we obtain
\begin{equation}
\la S_i^{\pm}\ra=\frac{H_i^{\pm}}{\tilde{H_i^z}}\;
\la S_i^z\ra\;. \label{35} \end{equation}
With this regularity condition, we complete the proof that ${\bf L}^0$
is a left eigenvector of ${\bf\Gamma}$ with eigenvalue zero since we
obtain for the non-diagonal elements \\ \\
\hspace*{0.5cm} see equation (\ref{36}) above. \\ \\
Hence, $N$ out of the $3N$ eigenvalues of the multilayer matrix 
${\bf\Gamma}$ must be zero. \addtocounter{equation}{1}

Apart from dimension, the equations determining the correlation
functions have the same form as for the monolayer case:
\begin{equation}
{\bf C}^{mn}={\bf R\;{\cal L}\;\bf A}_{\eta=-1}^{mn}+{\bf R}^0\;{\bf L}^0
\;{\bf C}^{mn}\;. \label{37} \end{equation}
The matrices ${\bf R}$ and ${\bf {\cal L}}$ have to be constructed
from the right and left eigenvectors corresponding to non-zero
eigenvalues as before, whereas the matrices ${\bf R}^0$ and ${\bf
L}^0$ are constructed from the $N$ eigenvectors with eigenvalues zero.

In order to compute the matrix ${\bf\Gamma}$ when iterating
equations (\ref{37}), one has to solve the linear system
of equations (\ref{35}), 
\begin{eqnarray}
Z_i\la S_i^\pm\ra && \hspace{-0.4cm} -J_{i,i+1}\,\la S_{i+1}^{\pm}\ra\,
\la S_i^z\ra-J_{i,i-1}\,\la S_{i-1}^{\pm}\ra\,\la S_i^z\ra \nonumber \\
&=&B^{\pm}\,\la S_i^z\ra\;; \hspace{0.5cm} i=1,\ldots,N\;,
\label{38} \end{eqnarray}
in each iteration step.  For general spin $S$, one can express all
higher correlations occuring in equation (\ref{37}) in terms of the
moments of $\la(S^z_i)^{2S+1}\ra$ for layer $i$ by using all
regularity conditions with $m+n\leq2S$. Again the largest moment can
be expressed by lower ones using $\prod_{M_S} (S_i^z-M_S)=0$.

\subsection{The effective anisotropy}
If, in addition to the orientation angle, one is interested in the effective
(temperature-dependent) anisotropy coefficient $K_2(T)$, a quantity which is
accessible in experiment, one needs a working expression for the free energy.
For a derivation of this expression, which is also used by experimentalists to
extract $K_2(T)$, we refer to the books of Landau and Lifschitz \cite{LaLi} and
of Vonsovskii \cite{Von74}. To lowest order the free energy reads
\begin{eqnarray}
F(T) & = & \sum_{i=1}^N\; F_i(T) \nonumber \\
F_i(T) & = &
-\frac{1}{2}\sum_{<l>}J_{il}\,{\bf S}_i\cdot{\bf  S}_l
-K_{2,i}(T)\,\cos^2\Theta_i-{\bf B}\cdot{\bf S}_i \nonumber \\
& &+\frac{1}{2}\sum_{l}\frac{g_{il}}{r^5_{il}}\Big({\bf S}_i\cdot{\bf S}_l
-3\,({\bf S}_i\cdot{\bf r}_{il})\,({\bf S}_l\cdot{\bf r}_{il})\Big)\;.
\label{41b} \end{eqnarray}

As in reference \cite{FJK00}, the temperature-dependent anisotr\-opy 
$K_{2,i}(T)$ for each layer $i$ is calculated non-perturbatively
by minimizing the free energy with respect to the layer-dependent 
reorientation angle $\Theta_i$. From the condition \break 
$\partial F(T)/\partial\Theta_i(\Theta_{0,i})=0$, we find with 
${\bf B}=(B^x,0,B^z)$
\begin{eqnarray}
& &K_{2,i}(T)=\frac{M_i(T)}{2\sin\Theta_{0,i}\cos\Theta_{0,i}}\Big[
\nonumber \\
& & \cos\Theta_{0,i}\Big(B^x+J_{i,i+1}\,M_{i+1}(T)\,\sin\Theta_{0,i+1}
\nonumber \\ 
&& +J_{i,i-1}\,M_{i-1}(T)\,\sin\Theta_{0,i-1}+T_i^{\sin}\Big) \nonumber \\
&& -\sin\Theta_{0,i}\Big(B^z+J_{i,i+1}\,M_{i+1}(T)\,\cos\Theta_{0,i+1}
\nonumber \\ 
&& +J_{i,i-1}\,M_{i-1}(T)\,\cos\Theta_{0,i-1}-2\,T_i^{\cos}\Big)\Big] \;, 
\label{42} \end{eqnarray}
where the magnetization $M_i(T)=\sqrt{\la S_i^x\ra^2+\la S_i^z\ra^2}$,
and the equilibrium polar angle $\Theta_{0,i}=\arctan(\la S_i^x\ra/\la
S_i^z\ra)$ are determined from the magnetization components 
$\la S_i^x\ra$ and $\la S_i^z\ra$ calculated from the Green's function
method.  $J_{i,i\pm 1}$ is the exchange interaction between
neighboring layers, and
\begin{equation}
T_i^{\left\{\sin \atop \cos\right\} } = \sum_{j=1}^N\;
g_{i,j} M_{j} \left\{\sin\Theta_{0,j} \atop \cos\Theta_{0,j}\right\}\;
T^{|i-j|}\;, \end{equation}
$T^{|i-j|}$ being the dipole lattice sum occuring in equation
(\ref{39}). For a single layer $N=1$, we obtain  equation (31) of
reference \cite{FJK00}, with an additional term
$(3/2)\,g_{11}\,M_1^2(T)\;T^0$ due to the dipole coupling. The total
effective anisotropy $K_2(T)$  of the thin film is given by
\begin{equation}
K_2(T)=\sum_{i=1}^N K_{2,i}(T)\;. \label{45} \end{equation} 
This procedure for the determination of the effective anis\-otropies
$K_{2,i}(T)$ is non-perturbative in the sense that the magnetization
and the orientation angle in equation (\ref{42}) are calculated from
the {\it full} Hamiltonian, in contrast to a thermodynamic
perturbation theory where one splits the Hamiltonian into two terms,
e.g.\ \cite{EFJK99}.  In the numerical calculations, we will normalise
the anisotropy coefficient in the Hamiltonian to $K_{2,i}/S(S-1/2)$,
in order to guarantee that $K_{2,i}(T)/K_{2,i}(T=0)=1$ for
$T\rightarrow 0$.

We are aware of the fact that using only $K_2$ in the Hamiltonian can
lead to an effective $K_4(T)$ \cite{theo}, which, although it turns
out to be very small in an analysis within mean field theory, has some
effect on the nature of the phase transition and on the phase
diagram. We do not  try to extract a corresponding $K_4$ term here,
because we do not calculate the order of the reorientation phase
transition or a phase diagram in the present paper.
\begin{figure}[t] \label{Fig1}
\includegraphics[width=4.5cm,height=8.5cm,angle=-90]{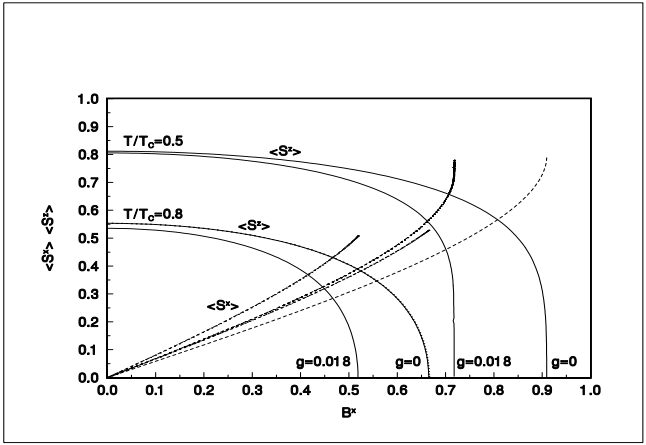} 
\vspace{2cm}

\caption{ The components of the magnetization $\la S^z\ra$ (solid
lines) and $\la S^x\ra$ (dashed lines) of a spin $S=1$ monolayer are
shown  as functions of the external magnetic field in $x$-direction,
$B^x$, without ($g=0$) and with dipole coupling ($g=0.018$, the value
estimated for Ni), and $J=100$ and $K_2=1$. The dipole coupling
renormalizes the external magnetic field.  Two different reduced
temperatures ($T/T_C=0.5$ and $0.8$) are considered, where $T_C$ is
the Curie temperature for perpendicular magnetization. The 
$\la S^x\ra$ components are plotted only up to the temperature where 
$\la S^z\ra \rightarrow 0$.}
\end{figure}

\section{Results}
In this section we show results of the calculations described above.
First we discuss the effect of the dipole coupling on the
reorientation of the magnetization for a monolayer with spin $S=1$.
Secondly, we discuss the case of a single layer with spins
$S>1$. Thirdly, we treat ferromagnetic films consisting of $N$ layers.

\subsection{The effect of the dipole coupling on the magnetic
reorientation of a monolayer}

Since in reference  \cite{FJK00} the dipole coupling has not been
taken into account explicitly, we investigate in this subsection the
action of this interaction on the magnetic reorientation in the case
of a monolayer with spin quantum number $S=1$.  As discussed above, the
exchange coupling is treated by RPA, the single-ion anisotropy terms
according to the Anderson-Callen decoupling, and the dipole coupling
is considered within the simplified (non-dispersive mean field)
approximation described in Appendix A.  We use the parameters $J=100,
K_2=1$ chosen in reference \cite{FJK00}. The dipole coupling strength
is set equal to $g_{11}\equiv g=0.018$ or $g=0.066$, which refers to
the cases of Ni or Co by calculating the relative strength of $J/g$,
where $J$ is estimated from the corresponding bulk Curie
temperatures. The external magnetic field is directed along the
$x$-axis, ${\bf B}=(B^x,0,0)$.
\begin{figure}[t] \label{Fig2}
\includegraphics[width=4.5cm,height=8.5cm,angle=-90]{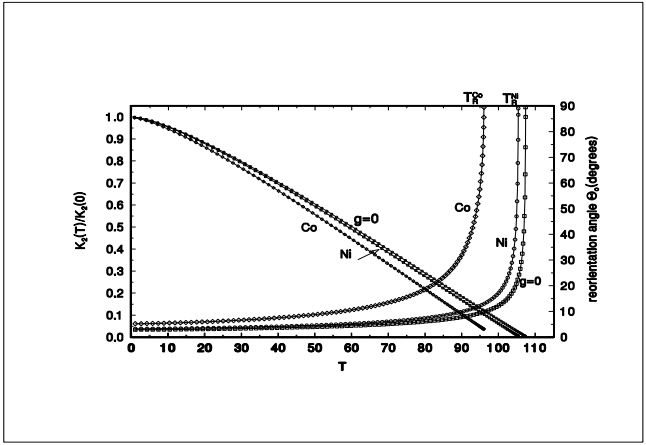} 
\vspace{2cm}

\caption{ Effective anisotropy $K_2(T)/K_2(0)$ and equilibrium
reorientation angle $\Theta_0$ of a spin $S=1$ monolayer are shown as
functions of the temperature $T$ without ($g=0$) and with dipole
coupling ($g=0.018$ for Ni and $g=0.066$ for Co). A magnetic field
$B^x=0.1$ is applied, and $J=100$ and $K_2=1$.  }\end{figure}

In Figure 1 we plot the components of the magnetization $\la S^z\ra$
and $\la S^x\ra$ without ($g=0$) and with ($g=0.018$) dipole coupling
for the temperatures $T/T_C=0.5$ and $T/T_C=0.8$ as a function of the
transverse field $B^x$. Here $T_C$ is the Curie temperature calculated
for a perpendicular magnetization. As expected from equations
(\ref{8},\ref{39}), the dipole coupling diminishes the action of the
uniaxial anisotropy and the magnetic field in the $z$-direction,
leading to a reduction of $\la S^z\ra$, and enhances the action of the
transverse components; consequently, $\la S^x\ra$
increases. Therefore, the reorientation field $B_R^x$, at which 
$\la S^z\ra$ vanishes, becomes smaller for increasing dipole coupling
\break strength.  

In Figure 2 we plot the equilibrium reorientation angle
$\Theta_0(T)=\arctan(\la S^x\ra/\la S^z\ra)$ for
$B^x=0.1$ as a function of the temperature $T$. The following dipole
coupling strengths are considered: $g=0$, $g=0.018$ (estimated for
Ni), and $g=0.066$ (estimated for Co). With increasing dipole coupling
strength the reorientation temperature $T_R$, at which $\la S^z\ra$
vanishes ($\Theta_0=90^\circ$), decreases. The corresponding effective
(temperature-dependent) anisotropy $K_2(T)/K_2(0)$ as obtained from
equation (\ref{42}), is also shown.  Since the results shown in Figure
2 are obtained for a finite magnetic field $B^x=0.1$, $\la S^x\ra$
remains finite for $T\ge T_R$, and $K_2(T)$ does not vanish completely
at $T_R$. For the  coupling constants under consideration the overall
behaviour of $K_2(T)/K_2(0)$ does not change for different $g$. 
\begin{figure}[t] \label{Fig3}
\vspace*{-0.5cm} 

\hspace*{-0.5cm} \includegraphics[width=7.5cm,height=10cm,
angle=-90,bb=120 100 550 750,clip]{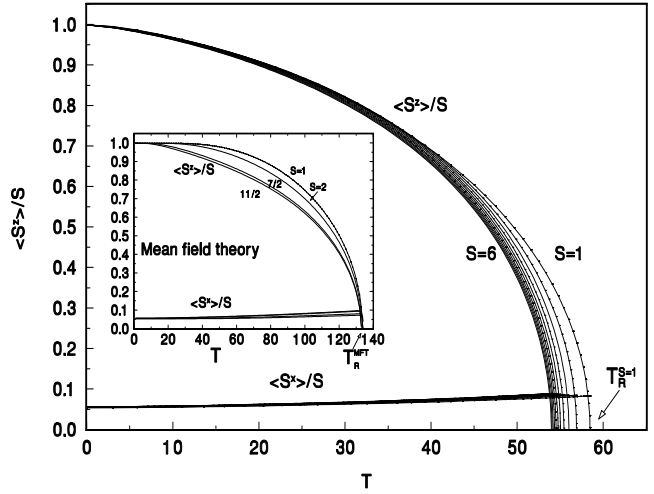} \vspace{0cm}

\caption{ Normalized magnetizations $\la S^z\ra/S$ and $\la S^x\ra/S$ 
for a monolayer as functions of the temperature for all integral
and half-integral values of the spin between $S=1$ and $S=6$
calculated with the Green's function theory. The reorientation
temperature $T_R^S$ depends slightly on $S$. The inset shows
corresponding results where all interactions are treated with MFT for
spins $S=1,2,7/2,$ and $11/2$. In this case, the reorientation
temperature $T_R^{MFT}$ does not depend on $S$.  All parameters,
$B^x=0.1$, $J=100$, $K_2=1$, and $g=0.018$, are scaled as discussed in
the text.  }\end{figure}

\subsection{The monolayer for ${\bf S>1}$}

In this subsection we investigate the effect of different spin quantum
numbers $S$ on the magnetic reorientation. We consider a monolayer
with the interaction parameters  used for the results of Figure 1.  In
order to compare the results for different $S$ we have scaled these
parameters in the following way: $J\rightarrow J/S(S+1)$,
$B\rightarrow B/S$, $g\to g/S(S+1)$, and $K_2=K_2/S(S-1/2)$. The
scaling of $K_2$ guarantees the property $\lim_{T\rightarrow
0}K_2(T)/K_2(0)=1$ for $B_x\rightarrow 0$.

In Figure 3 we display the normalized magnetizations $\la S^z\ra/S$ as
functions of the temperature $T$ for all integral and half-integral
spins ranging between $S=1$ and $S=6$. The reorientation temperature,
$T_R(S)$, becomes smaller with increasing $S$ but at a rapidly
decreasing rate as $S$ increases.  The small external magnetic field
in $x$-direction ($B^x=0.1$) induces a finite $x$-component of the
magnetization  $\la S^x\ra/S$ for $T=0$, which increases slightly with
increasing temperature.  These results are compared with results of
calculations where a mean field approximation (MFT) is performed for
all interactions \cite{JeB98}, see the inset of Figure 3. Within this
aproximation a more pronounced spin dependence of the magnetization
curves is observed.  On the other hand, due to the scaling of the
coupling parameters the mean field reorientation temperatures
$T_R^{MFT}(S)$ are independent of the spin
value $S$. Note, however, that $T_R^{MFT}$ is more than a factor of
two larger than the reorientation temperature as calculated from the 
Green's function theory - this is due to missing correlations in MFT.

In Appendix B it is shown that the correlations ocurring in the
equations of motion can be expressed by higher moments of the
magnetization, which are determined by the regularity conditions. As
an example, we present in Figure 4 results of the spin wave
calculation for the normalized moments $\la (S^z)^n\ra/S^n$ for spin
$S=11/2$ and for $n=1,2,\ldots,11$ as functions of the
temperature. The odd moments approach zero for $T\rightarrow T_R$,
whereas the even moments approach a finite value for $T\rightarrow
T_R$ as expected physically.  For $n=2$ one obtains for example
$\la(S^z)^2\ra/S^2\rightarrow S(S+1)/3S^2$.
\begin{figure}[t] \label{Fig4}
\includegraphics[width=4.5cm,height=8.5cm,angle=-90]{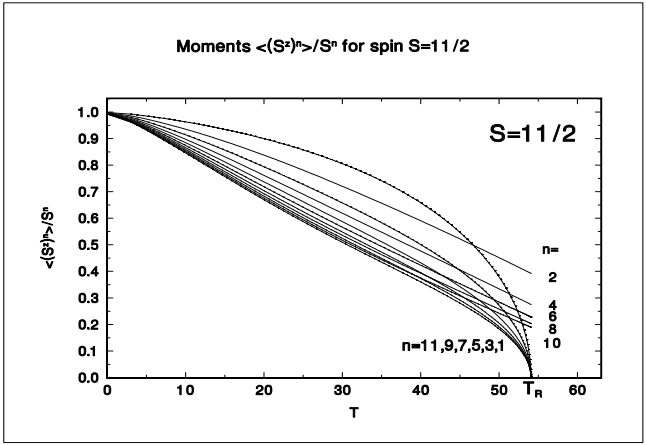} 
\vspace{2cm}

\caption{ Results of a Green's function calculation for the normalized
moments $\la(S^z)^n\ra/S^n$ for a monolayer and spin $S=11/2$ for
$n=1,2,\ldots,11$. $T_R$ is the reorientation temperature.}\end{figure}

In Figure 5 the equilibrium reorientation angles $\Theta_0$, and in
Figure 6 the corresponding effective anisotropy coefficients
$K_2(T)/K_2(0)$ are shown as functions of the temperature for spin
quantum numbers ranging from $S=1$ to $S=6$, as calculated from the
Green's function method.  The temperature dependence of both these
quantites does not vary markedly with $S$. In the insets of the
figures are the corresponding results for $\Theta_0$ and
$K_2(T)/K_2(0)$ as obtained from a MFT for all interactions. As
already seen for the magnetizations, a more pronounced spin dependence
of $K_2(T)/K_2(0)$ is observed here also for MFT. Evidently, the spin
quantum number $S$ has a larger influence on single-spin excitations
(MFT) than on collective magnetic excitations (spin waves).

If scaled coupling constants are used, only a weak dependence of the
spin quantum number $S$ on the magnetic quantities
such as the magnetization, the reorientation angle, and the effective
anisotropy is observed within the Green's function method.
Thus, one may perform calculations with a low spin, for
which a considerably smaller system of equations has to be solved
self-consistently. Results for higher spins can then be obtained by scaling.
This is less justified within the MFT approximation.

We stress a result already obtained in reference \cite{FJK00} for
spin $S=1$, namely that the effective anisotropies $K_2(T)$  as
calculated within the RPA and within  MFT have different temperature
behaviours, particularly at low temperatures. In this temperature
regime, the MFT exhibits an exponential behaviour, and RPA an almost
linear behaviour of $K_2(T)$.  Consequently, the use of MFT would lead
to a considerably smaller value for $K_2(0)$ than that obtained with
RPA, when the observed values of $K_2(T)$ (measured typically at
$T/T_C\simeq 0.7$, e.g.\ \cite{exp}) are extrapolated to $T=0$. Note
that only at $T=0$ are anisotropy coefficients available from
ab-initio calculations, e.g.\ reference \cite{HB97}. These theoretical
values can only be compared with experiment by extrapolating
measurements at finite temperatures down to zero with the help of a
theoretical model. Because of the drawbacks of MFT we propose
performing this extrapolation with the results from the Green's function
method.
\begin{figure}[t] \label{Fig5} \vspace*{-0.5cm}

\includegraphics[width=7.5cm,height=9cm,angle=-90,,bb=150 90 600
700,clip]{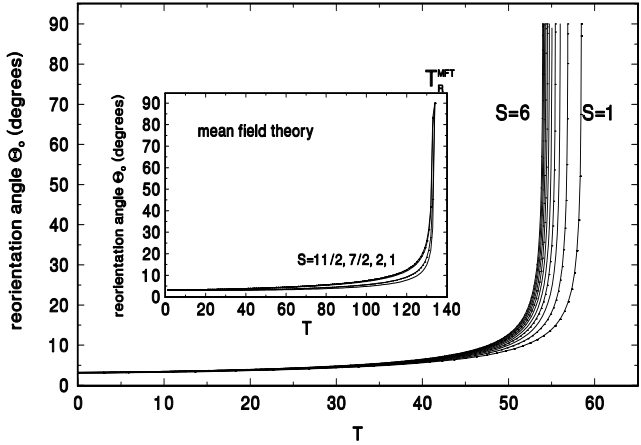} \vspace{-1cm}

\caption{Equilibrium reorientation angle $\Theta_0$ as a function of the
temperature for the systems of Figure 3 calculated with the Green's 
function theory. The inset shows the corresponding results when applying 
MFT to all interactions. }\end{figure}

\begin{figure}[t] \label{Fig6}
\includegraphics[width=8cm,height=9.2cm,angle=-90,,bb=150 90 600
700,clip]{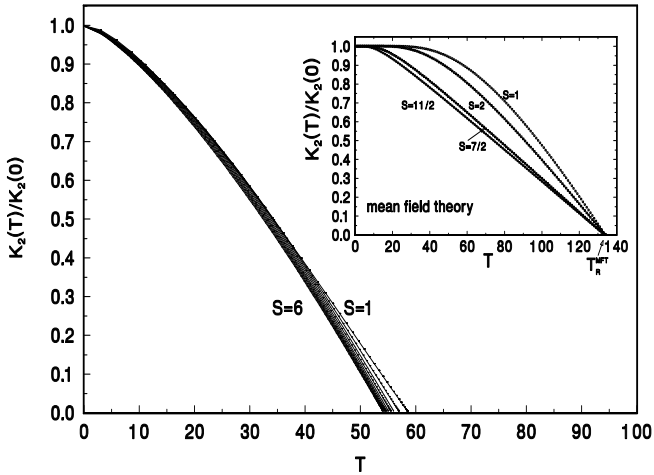} \vspace{-1cm}

\caption{ Effective anisotropy $K_2(T)/K_2(0)$  calculated with the
Green's function method as a function of the temperature for the
systems of Figure 3.  The inset shows the corresponding MFT results.
}\end{figure}

\subsection{Ferromagnetic films with N layers}
In this subsection we demonstrate that the formalism developed in
Section 2 can be applied to the case of many layers. We study the
magnetization, the reorientation angle, and the effective anisotropy
as functions of the film thickness (characterized by the number of
atomic layers, $N$). In all the examples shown in this subsection, the
dipole coupling is included within the simplified mean field treatment
as discussed above.

As examples we treat simple cubic films consisting of $N$ layers with
spin $S=1$, using the same coupling parameters for all film layers
(homogeneous film): $J=J_{ik}=100$, $K_{2,k}=K_2=1$, $g_{ik}=g=0.018$.
These couplings are scaled as in Section  3.2 : $J\to J/S(S+1)$, $g\to
g/S(S+1)$, $K_2\to K_2/S(S-1/2)$, and $B\to B/S$. All quantities are
calculated for a small external magnetic field ${\bf B}=(0.1,0,0)$.

In Figure 7 we show the sublayer magnetizations $\la S_i^z\ra$,
$i=1,\ldots,N$, as functions of the temperature for film thicknesses
ranging between $N=1$ and $N=19$ layers. As expected, and also seen in
MFT calculations \cite{JeB98,theo}, for a homogeneous film, the
magnetization of the surface layers is smaller than those of the
interior layers because of the smaller coordination number for the
surface. Also, the magnetizations of the $i$th and the $(N-i+1)$th
layer are the same (twofold symmetry).  For the parameters under
consideration the reorientation temperature $T_R$ is close to the
Curie temperature for perpendicular magnetization. Therefore, we see
in Figure 7 that the value for the reorientation temperature $T_R$
(defined as the temperature where $\la S^z_i\ra=0$) exhibits a
saturation behaviour as a function of the film thickness.  Whereas
there is a steep rise from $T_R(N=1)=58.55$ for the monolayer to
$T_R(N=3)=103.08$ for the trilayer, there is only a small difference
between $T_R(N=17)=135.11$ and $T_R(N=19)=135.67$. 
\begin{figure}[t] \label{Fig7}
\includegraphics[width=6.5cm,height=10.5cm,angle=-90]{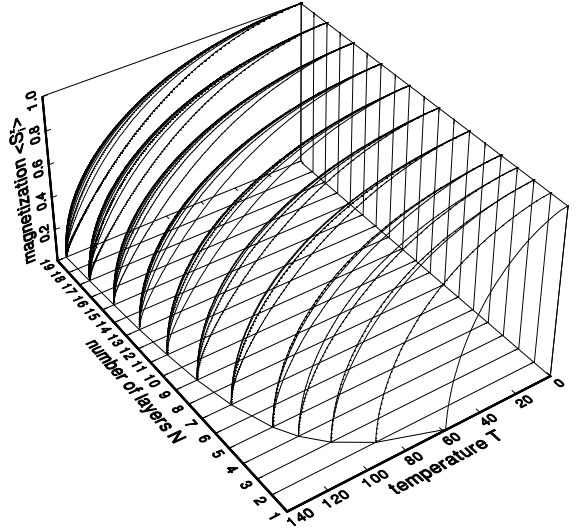} 
\vspace{2cm}

\caption{ Sublayer magnetizations $\la S^z_i\ra$ as functions of the
temperature for thin ferromagnetic films with $N$ layers and spin
$S=1$. The reorientation temperatures $T_R^N$ for the different films
can be read off from the curve in the $N-T$ plane, where $\la S^z_i\ra
=0$.  The same parameters are used for all layers: $B^x=0.1$, $J=100$,
$K_2=1$, and $g=0.018$.  They are scaled as described in the text.
}\end{figure}

Because in experiment only the average orientation of the thin film 
magnetization is measured one has to calculate this quantity from the model.
In Figure 8 we show the average equilibrium reorientation angles
$\Theta_0(N,T)$ of  thin films with different thicknesses $N$ as  functions
of the temperature, where
\begin{equation}
\Theta_0(N,T)=\arctan\;\frac{\frac{1}{N}\sum_{i=1}^N \la S_i^x\ra}
{\frac{1}{N}\sum_{i=1}^N \la S_i^z\ra} \;. \label{44}
\end{equation}
An alternative to calculating the average orientation angle is first to
calculate the angles for each layer,
$\Theta_{0,i}=\arctan(\la S_i^x\ra/\la S_i^z\ra)$, and then to
average over the angles. The difference between both procedures turns 
out to be tiny. 
\begin{figure}[t] \label{Fig8}
\includegraphics[width=4.5cm,height=8.5cm,angle=-90]{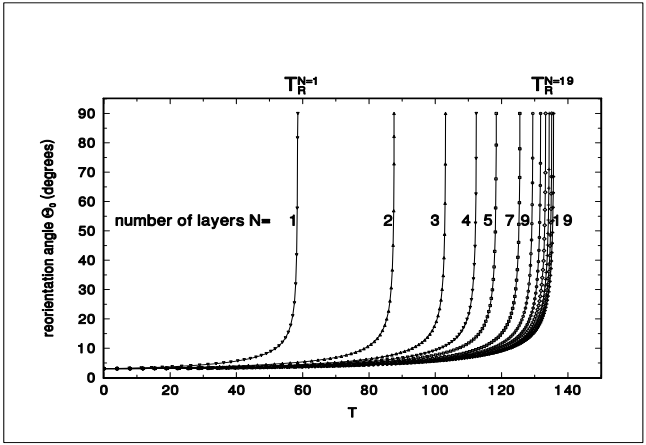} 
\vspace{2cm}

\caption{ The average equilibrium reorientation angle $\Theta_0$ as a
function of the temperature and film thickness.  $N$ is the number of
layers in each film and $T^N_R$ are the reorientation temperatures
($\Theta_0=90^\circ$).  }\end{figure}

In Figure 9 the average effective anisotropies $K_2(N,T)$ $/K_2(N,0)$ of
thin films with different thicknesses $N$,  calculated from equation
(\ref{45}), are shown as  functions of the temperature.
With increasing film thickness the action of the effective anisotropy
extends to higher temperatures.  In the inset we also show for $N=1$ and
$N=19$ that the dependence of the anisotropies on the temperature is
somewhat different for different layer thicknesses if one scales the
temperature to the respective reorientation temperatures:
$T/T_R(N)$. The curvature of the curve for the thick film ($N=19$) is
more pronounced as that for the monolayer.
\begin{figure}[t] \label{Fig9} \vspace*{-0.5cm}

\includegraphics[width=7.5cm,height=9cm,angle=-90,bb=150 90 600
700,clip]{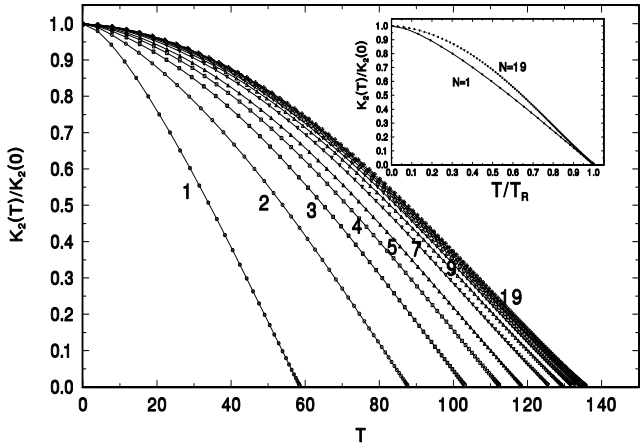} \vspace{-1cm}

\caption{ Average effective  anisotropy $K_2(T)/K_2(0)$ as a function
of the temperature and film thickness $N$.  The inset demonstrates the
different functional dependence of $K_2(T)/K_2(0)$ on the temperature
for layers with $N=1$ and $N=19$ if the temperature $T$ is scaled with
respect to the reorientation temperature $T_R$.  }\end{figure}

\section{Discussion and conclusions}
In the present paper we have extended the model of reference
\cite{FJK00} in various respects and we have developed a succinct
formulation of the final equations. This was made possible by
utilizing the eigenvectors as well as the eigenvalues of the matrix
governing the equations of motion for the set of Green's functions,
which has to be introduced when the calculation of several
non-vanishing components of the magnetization is required. This new
procedure, which for the monolayer and spin $S=1$ is fully equivalent
to our earlier treatment \cite{FJK00}, provides a practicable way of
extending the Green's function spin wave theory to the reorientation
of the magnetization of ferromagnetic films consisting of {\it many}
layers and for general spin $S$.

We have applied the new method to the monolayer case with spins $S\geq
1$.  We have found that the spin dependence of the magnetizations and
the anisotropies as functions of the temperature is considerably less
pronounced in RPA than in MFT if a proper scaling of the parameters of
the model is applied . The corresponding curves saturate much more
quickly in RPA than in MFT with increasing spin quantum number
$S$. The temperature for complete reorientation $T_R$, on the other
hand, does not change with $S$ in MFT, whereas there remains a slight
spin dependence in RPA.

For the monolayer with spin $S=1$, we have investigated in detail the
influence of the dipole coupling on the reorientation
problem. Because, for realistic dipole coupling strengths, we found no
big differences in treating the dipole coupling with its RPA or
(non-dispersive) MFT approximations , cf.\ Appendix A, we chose to
include the dipole coupling by means of the latter, which is
relatively simple to handle and requires only a renormalization of the
external magnetic field.

We emphasize that only by using our new method have we been able to
treat the magnetic reorientation within a Green's function approach
for films with several layers.  We have studied the field-induced
magnetic reorientation and the effective anisotropy as functions of
the film thickness and temperature for spin $S=1$ films.
Investigations of films with $S\geq 1$ present no problem; they are
only more time consuming.

In the present paper, we have only studied homogeneous films, but the
method can also treat inhomogeneous films or multilayers by using
layer-dependent coupling constants and magnetic moments. The magnetic
reorientation could then be calculated for thin film or multilayer
systems investigated experimentally.  This will be pursued in
forthcoming studies.

A few words concerning the applications of the present model are in
order. A prerequisite is that the investigated systems can be modelled
in a reasonable way by a local spin model of Heisenberg
type. Moreover, it is required that higher order anisotropies are not
important, a condition which is often not fulfilled.

Most results of the present paper are obtained for a small transverse
field which initializes the reorientation. It is certainly of interest
to calculate also phase diagrams in which the magnetic field is
varied, and to compare with results obtained in reference \cite{Mil97}
within a schematic model and in reference \cite{Us2000} on the basis
of mean field theory.

The possibility of treating spins $S>1$ would also allow the treatment
of the fourth-order uniaxial anisotropy $K_4$ as well as the quartic
in-plane anisotropy. This, however, requires a proper decoupling
procedure for the corresponding terms in the Green's function theory
which we do not have available at the moment.  A phase diagram in the
$K_2-K_4$ plane would then show the region of stable magnetization
directions and would help in exploring the locations of temperature or
film thickness driven magnetic reorientations.  Also, one could
determine whether the magnetic reorientation happens continuously or
discontinuously.

\section*{Appendix A: The treatment of the dipole coupling}

In this Appendix we show how the long-range magnetic dipole coupling
can be considered within the RPA treatment of the magnetic
reorientation. Furthermore, we show that a simplified treatment for
interaction strengths small compared to the exchange interaction leads
to a satisfactory description of the magnetic properties. For
simplicity we consider only the case of a single layer ($N=1$).

As it should be, the magnetic dipole interaction leads to an
additional dispersion in the magnon dispersion relation $\omega({\bf
k})$.  Applying the generalized Tyablikov (RPA) decoupling, equation
(\ref{4a}), to the dipole interaction in the Green's function
equations of motion (\ref{3}), $\la\la[S_i^{\alpha},{\cal H}^{\rm
dipole}]_-;$ $(S_j^z)^m(S_j^-)^n\ra\ra$, one obtains the following
additional terms on the left side of equations (\ref{6}) 
\begin{equation}
\left( \begin{array}{ccc}
-T_{\bf k}^+     & -T_{\bf k}^-          & -T_{\bf k}^z \\
(T_{\bf k}^-)^*  & T_{\bf k}^+           & (T_{\bf k}^z)^* \\
T_{\bf k}^{z\pm} & -(T_{\bf k}^{z\pm})^* & 0
\end{array} \right)
\left( \begin{array}{c} G_{\eta}^{+,mn}\\ G_{\eta}^{-,mn}\\ G_{\eta}^{z,mn}
\end{array} \right)\;. \label{B1} \end{equation}
Here a 2D Fourier transformation has been applied, and
\begin{eqnarray}
T_{\bf k}^+&=&g\;\la S^z\ra\Big(T_{20}^0+T_{02}^0+\frac{1}{2}T_{20}^
{\bf k}+\frac{1}{2}T_{02}^{\bf k} \Big) \nonumber \\
T_{\bf k}^-&=&\frac{3}{2}\;g\;\la S^z\ra
(T_{20}^{\bf k}-T_{02}^{\bf k}+2i\;T_{11}^{\bf k} )
\label{B2} \\
T_{\bf k}^z&=&g\;\la S^+\ra \Big(T_{20}^{\bf k}
+T_{02}^{\bf k}+\frac{1}{2}T_{20}^{0}+\frac{1}{2}T_{02}^{0} \Big) \nonumber\\
T_{\bf k}^{z\pm}&=&\frac{g}{4}\bigg( \la S^-\ra
(T_{20}^{\bf k}+T_{02}^{\bf k}-T_{20}^{0}-T_{02}^{0}) \nonumber\\
&& -3\la S^+\ra(T_{20}^{\bf k}-T_{02}^{\bf k}-2i\;T_{11}^{\bf k}) \bigg)\;,
\nonumber \end{eqnarray}
where
\begin{equation}
T_{\mu\nu}^{\bf k}=\sum_{lm}\frac{(x_l)^\mu(y_m)^\nu}
{(x_l^2+y_m^2)^{5/2}}\;\exp(ik_xx_l)\;\exp(ik_yy_m)
\label{B3} \end{equation}
are oscillating lattice sums, which can be evaluated with Ewald
summation techniques as outlined for instance in reference \cite{jens97}.

As mentioned in the main body of the paper, the RPA treatment of the
magnetic dipole coupling complicates the calculation of the
magnetization considerably because of the presence of complex terms
and dispersive (${\bf k}-$depen\-dent) terms.  Thus, as an
approximation we neglect now the dispersive parts in the equations of
motion (\ref{B1}) coming from the dipole coupling, and retain the
non-dispersive terms only. This corresponds to a mean field treatment
of the dipole coupling. Then equations (\ref{B2}) reduce to
\begin{eqnarray}
T_{\bf k}^+&=&g\;\la S^z\ra(T_{20}^0+T_{02}^0) \nonumber\\
T_{\bf k}^-&=&0  \\
T_{\bf k}^z&=&\frac{g}{2}\;\la S^+\ra(T_{20}^{0}+T_{02}^{0}) \nonumber\\
T_{\bf k}^{z\pm}&=&-\frac{g}{4}\;\la S^-\ra(T_{20}^{0}+T_{02}^{0})\;.
\nonumber \label{B4} \end{eqnarray}
This simplification allows the dipole coupling to be
taken into account by a  renormalization of the external
magnetic field, and leads to equations (\ref{39}) of Section 2.

By comparing to RPA, we shall show now that this approximation leads
to satisfactory results for small dipole coupling strengths as found
for instance in ferromagnetic $3d$- transition metal thin films.
Because the general RPA treatment of the dipole coupling turns out to
be fairly complicated, we consider only two limiting cases which are
manageable, a perpendicular and an in-plane magnetization.  An
external magnetic field is not considered but can be easily added.

For spin $S=1$, one needs the Green's functions $G_{ij}^{\pm,mn}$
$=\la\la S_i^\pm;(S_j^z)^m(S_j^-)^n\ra\ra$ for $n=1$ and for $m=0$ and
$m=1$. In case of the perpendicular magnetization, use of the RPA
decoupling for the exchange interaction and the dipole coupling, and
the Anderson-Callen decoupling for the single-ion anisotropy
($\Phi=\la S^z\ra\la S^zS^z\ra$ for $S=1$) leads to the following
equations of motion (the $z$-axis is directed perpendicular to the
plane) 
\begin{equation}
\left(\begin{array}{cc}\omega-a & b \\ -b^* & \omega+a
\end{array}\right)
\left(\begin{array}{c} G^{+,m1}({\bf k},\omega) \\
G^{-,m1}({\bf k},\omega) \end {array} \right) =
\left(\begin{array}{c} A^{+,m1} \\ A^{-,m1} \end{array} \right) \;,
\label{B5} \end{equation} 
with 
\begin{eqnarray}
a&=&\la S^z\ra\bigg(J\,(q-\gamma_{\bf k})-g\Big(T_{20}^0+T_{02}^0
+\frac{1}{2}T_{20}^{\bf k}+\frac{1}{2}T_{02}^{\bf k}\Big) \nonumber \\
&& -K_2\;\la S^zS^z\ra \bigg) \nonumber \\
b&=&\frac{3}{2}\;g\;\la S^z\ra \Big(T_{20}^{\bf k}-T_{02}^{\bf k}
+2i\,T_{11}^{\bf k} \Big) \;. \label{B6} \end{eqnarray} 
Solving these equations for the Green's functions and applying the spectral
theorem we obtain the following correlation functions
\begin{eqnarray}
\la(S^z)^mS^-S^+\ra&=&-\frac{1}{2}\,A_{-1}^{+,m1} \nonumber \\
&& \hspace*{-2cm} +\frac{1}{2\epsilon}\;
\Big(a\,A_{-1}^{+,m1}-b\,A_{-1}^{-,m1}\Big)
\coth\Big(\frac{\beta\epsilon}{2}\Big) \;, \nonumber\\
\la(S^z)^mS^-S^-\ra&=&\frac{1}{2}\,A_{-1}^{-,m1} \nonumber \\ 
&& \hspace*{-2cm} -\frac{1}{2\epsilon}\;
\Big(a\,A_{-1}^{-,m1}-b^*\,A_{-1}^{+,m1}\Big)
\coth\Big(\frac{\beta\epsilon}{2}\Big) \;, \label{B7} \end{eqnarray} 
where the dispersion relation is 
\begin{equation}
\epsilon=\sqrt{a^2-|b|^2}\;\;. \end{equation} 
Using now for $S=1$ the identities $S^-S^+=2-S^z-(S^z)^2$ and
$(S^z)^3=S^z$, the inhomogeneities are given by 
\begin{eqnarray}
A_{-1}^{+,01}=2\la S^z\ra\;, & & A_{-1}^{+,11}=3\la (S^z)^2\ra
-\la S^z\ra-2\;, \nonumber \\ A_{-1}^{-,01}=0 \;, \hspace{1cm} & &
A_{-1}^{-,11}=\la S^-S^-\ra \;. \label{B8} \end{eqnarray} 
For $m=0$ we obtain 
\begin{eqnarray} 
\la(S^z)^2\ra&&=2-\la S^z\ra(1+2\phi_1)\;, \hspace{0.5cm}
\la S^-S^-\ra=2\la S^z\ra\phi_2^* \;, \nonumber \\ \label{B9}
\end{eqnarray}
and for $m=1$ 
\begin{eqnarray} 
\la S^z\ra-\la(S^z)^2\ra && =(3\la(S^z)^2\ra-\la
S^z\ra-2)\phi_1-\la S^-S^-\ra \phi_2 \;, \nonumber \\ \end{eqnarray}
where sums (integrals) over the first Brillouin zone have to be performed for
the quantities 
\begin{eqnarray}
\phi_1 &=& \frac{1}{N}\sum_{\bf k}\Big(\frac{a}{2\epsilon}\,
\coth\big(\frac{\beta\epsilon}{2}\big)-\frac{1}{2}\Big)\;,\nonumber \\ 
\phi_2 &=& \frac{1}{N}\sum_{\bf k}\frac{b}{2\epsilon}\,
\coth\big(\frac{\beta\epsilon}{2}\big) \;. \end{eqnarray}
From these equations we find the expectation values
\begin{eqnarray}
\la S^z\ra&=&\frac{1+2\phi_1}{1+3\phi_1+3\phi_1^2+|\phi_2|^2}\;, \\
\la(S^z)^2\ra &=& 2-\la S^z\ra(1+2\phi_1) \;.
\end{eqnarray}
These equations are solved self-consistently for $\la S^z\ra$
and $\la(S^z)^2\ra$.

In Figure 10 we show these expectation values with full RPA for the
dipole coupling ($g=0.018$ estimated for Ni) for a monolayer using the
parameters $J=100$, $K_2=1$, and for $g=0$ (no dipole coupling). For
the perpendicular magnetization, the $\la S^-S^-\ra$ correlation is
very small, and, within the line thickness of the curves, there is no
difference between the full RPA and a RPA where one puts $\la
S^-S^-\ra=0$ ($b=0$).  Also shown are the corresponding results for
the dipole coupling considered within the (non-dispersive) mean field
treatment. In this case, not only $b=0$ but also the {\bf k}-dependent
terms in equation (\ref{B6}) connected with the dipole coupling are
neglected, i.e.\ 
\begin{eqnarray}
a&=&\la S^z\ra\Big(J\,(q-\gamma_{\bf
k})-g\,(T_{20}^0+T_{02}^0)-K_2\,\la S^zS^z\ra \Big)\;,\nonumber\\
b&=&0\,. \label{B9a} \end{eqnarray}

As can be seen from Figure 10, the difference between a RPA and a MFT
treatment of the dipole coupling is small. 
\begin{figure}[t] \label{Fig10}
\includegraphics[width=4.8cm,height=9cm,angle=-90]{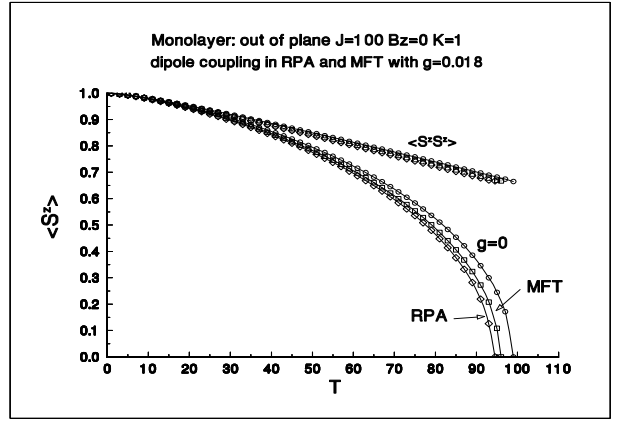} 
\vspace{2cm}

\caption{ Expectation value $\la S^z\ra$ for a monolayer with
perpendicular magnetization and for spin $S=1$ as a function of the
temperature. The dipole coupling ($g=0.018$ for Ni) is treated with
full  RPA, equation (\ref{B6}), and its mean field approximation,
equation (\ref{B9a}).  For comparison we show also the result with
vanishing dipole coupling ($g=0$). We have used $J=100$, $K_2=1$, 
and a vanishing external field ${\bf B}=0$.  }\end{figure}

\begin{figure}[t] \label{Fig11}
\vspace*{-0.2cm} 

\includegraphics[width=5cm,height=9.5cm,angle=-90]{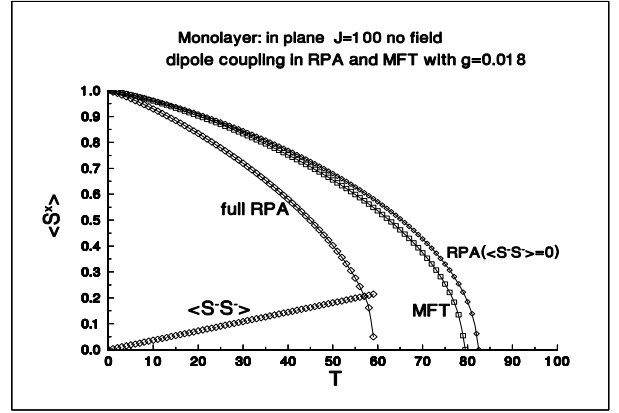} 
\vspace{2cm}

\caption{ In-plane magnetization $\la S^x\ra$ for a monolayer with
spin $S=1$ as function of the temperature. The dipole coupling
($g=0.018$) is treated with  full RPA, equation (\ref{B10}), with RPA
putting $\la S^-S^-\ra=0$, i.e.\ $b=0$ in equation (\ref{B9}), and with
MFT, equation (\ref{B11}). The $\la S^-S^-\ra_{RPA}$ correlation is
also shown.  }\end{figure}

The situation is somewhat different for an in-plane magnetization.  In
this case the single-ion anisotropy is not active ($\Phi=0$), and the
dipole term is, in accordance with the Mermin-Wagner theorem, the only
term which induces a finite magnetization.  In an RPA treatment of the
dipole coupling the magnetization in the plane is not isotropic and
one has to introduce an in-plane angle with respect to a main axis of
the square lattice. By putting this angle equal to zero one obtains
for the quantities $a$ and $b$, cf.\ equation (\ref{B5}), 
\begin{eqnarray}
a&=&\la S^z\ra\Big(J\,(q-\gamma_{\bf k})+\frac{g}{2}\,\big(T_{20}^0
+T_{02}^0-T_{02}^{\bf k}+2\,T_{20}^{\bf k}\big)\Big) \;, \nonumber \\ 
b&=&\frac{3}{2}\;g\;\la S^z\ra\;T_{02}^{\bf k} \;. \label{B10} 
\end{eqnarray} 
Using the mean field approximation for the dipole coupling by
neglecting the corresponding {\bf k}-dependent terms one obtains for the
in-plane magnetization
\begin{eqnarray}
a&=&\la S^z\ra\Big(J\,(q-\gamma_{\bf k})+\frac{g}{2}\,\big(
T_{20}^0+T_{02}^0\big)\Big) \;, \nonumber \\
b&=&0 \;. \label{B11} \end{eqnarray}

Within the formalism above, the quantization $z$-axis is now directed
in-plane, which corresponds to the $x$-direction of our original
reorientation problem.  The corresponding results are shown in Figure
11.  Here the difference between the full RPA and MFT is somewhat
larger, because the expectation value $\la S^-S^-\ra$ is not as
small as in the case for the perpendicular magnetization.  In
particular the Curie temperature is appreciably lower than in
MFT. This, however, is the most extreme case, and is an upper limit
for the error one makes if the dipole coupling is treated by MFT for
the reorientation problem.  As also shown in Figure 11 the MFT result
is close to the RPA result only when the $\la S^-S^-\ra$ correlation
is neglected.

In summary, the error is small for the perpendicular orientation and
will increase with increasing polar angle, but is expected to stay
below the error for the in-plane case when replacing the full RPA
treatment of the dipole coupling by the simplified, non-dispersive
MFT.  With this discussion in mind we use for the reorientation
problem of the present paper the mean field approximation for the
dipole coupling. We do so also because its full RPA treatment is quite
complicated and time-consuming.  If the ratio between the dipole
coupling strength and the exchange coupling becomes larger, as
expected e.g.\ for rare earth ferromagnets, the error will increase
and the simplified treatment is less justified. In this case one is
faced with the more complicated full RPA treatment of the dipole
coupling.

Finally, we mention that the magnetic dipole coupling can be taken
into account also via its mean field approximation when calculating
the magnetic reorientation  with the Schwinger-Boson theory
\cite{Timm00} . However, the proper inclusion  of the long-range
character of this interaction, which leads to additional
momentum-dependent terms in the magnon dispersion relation, is
practically impossible within this method.

\section*{Appendix B: Treating ${\bf S\geq 1}$}
In this Appendix we show how the regularity conditions, which have
been calculated in \cite{FJK00} for $S=1$, can be deduced for general
spin quantum numbers $S$.  The regularity condition (\ref{19}) for
$m=0,n=1$ yields
\begin{equation}
\frac{H^{\pm}}{\tilde H^z}=\frac{B^{\pm}}{Z}\;, \label{A1} \end{equation}
and can therefore be written for general $m,n$ in the form
\begin{equation}
-2\,Z\,A_{-1}^{z,mn}=A_{-1}^{+,mn}\,B^-+A_{-1}^{-,mn}\,B^+\;.
\label{A2} \end{equation}
The $z$-component of equation (\ref{23}), from which the correlations have
to be calculated, reads then explicitly
\begin{eqnarray}
& &2\,\frac{B^+B^-}{Z^2}\,\Big\la(S^z)^m(S^-)^nS^z\Big\ra
-\frac{B^-}{Z}\,\Big\la(S^z)^m(S^-)^nS^+\Big\ra \nonumber \\ 
&& \hspace*{2cm} -\frac{B^+}{Z}\,\Big\la(S^z)^m(S^-)^nS^-\Big\ra
\label{A33} \\
&=&\frac{1}{2}\,A_{-1}^{+,mn}\,\frac{E_{\bf k}}{\tilde{H^z}}\,
\frac{B^-}{Z}\,\Big[\frac{E_{\bf k}}{\tilde{H^z}}
-\coth\big(\frac{\beta E_{\bf k}}{2}\big)\Big] \nonumber \\ 
&& \hspace*{2cm} +\frac{1}{2}\,A_{-1}^{-,mn}\,
\frac{E_{\bf k}}{\tilde{H^z}}\,\frac{B^+}{Z}\,\Big[
\frac{E_{\bf k}}{\tilde{H^z}}+\coth\big(\frac{\beta E_{\bf k}}{2}\big)
\Big] \,. \nonumber \end{eqnarray}
We express all correlation functions occuring in this equation in a
standard form where all powers of $S^z$ are written to the left of the
powers of $S^-$:
\begin{equation}
C(m,n)=\la(S^z)^m(S^-)^n\ra\,.
\end{equation}
Then, with the relations $[S^z,(S^-)^n]_-=-n\,(S^-)^n$ and
$S^-S^+=S(S+1)-S^z-(S^z)^2$, we find that
\begin{eqnarray}
\la(S^z)^m(S^-)^nS^z\ra &=&n\,C(m,n)+C(m+1,n)\ ,\nonumber\\
\la(S^z)^m(S^-)^nS^+\ra&=&\big(S(S+1)-n\,(n-1)\big)\,C(m,n-1) \nonumber \\ 
&& \hspace*{-2.5cm} -(2n-1)\,C(m+1,n-1)-C(m+2,n-1)\ ,\nonumber\\
 \la(S^z)^m(S^-)^nS^-\ra&=&C(m,n+1)\;. \label{A5}
\end{eqnarray}
The commutators can also be expressed in terms of the
$C(m,n)$ using the binomial series
\begin{eqnarray}
A_{-1}^{z,mn}&=&-nC(m,n)\ ,\nonumber\\
A_{-1}^{+,mn}&=&\Big\la\Big[\big((S^z-1)^m-(S^z)^m\big)S^-S^+
+2\,S^z\,(S^z-1)^m \nonumber \\ 
&& \hspace*{-1cm} +(n-1)\,(n+2S^z)(S^z)^m\Big](S^-)^{n-1}\Big\ra \nonumber \\
&=&S(S+1)\sum_{i=1}^m \left( \begin{array}{c} m\\ i \end{array} \right)
(-1)^i\,C(m-i,n-1) \nonumber\\
&& \hspace*{-1cm} +\sum_{i=2}^{m+1} \left( \begin{array}{c} m+1\\ i 
\end{array} \right) (-1)^{i+1}\,C(m+2-i,n-1) \nonumber \\ 
&& \hspace*{-1cm} +(2n+m)\,C(m+1,n-1)+n(n-1)C(m,n-1)\;, \nonumber \\
A_{-1}^{-,mn}&=&\Big\la\big[(S^z+1)^m-(S^z)^m\big](S^-)^{n+1}\Big\ra 
\nonumber\\ &=&\sum_{i=1}^m \left( \begin{array}{c} m\\ i \end{array} 
\right)C(m-i,n+1) \;. \label{A6} \end{eqnarray}
Now by putting equation (\ref{A6}) into equation (\ref{A2}) the
regularity conditions for all $m$ and $n$ can be written in terms of
correlations defined in the standard form:
\begin{eqnarray}
&& 2\,Z\,n\,C(m,n)= \nonumber \\ 
&& B^-\Big[S(S+1)\,\sum_{i=1}^m\left(\begin{array}{c} m\\ 
i \end{array} \right)   (-1)^i\,C(m-i,n-1) \nonumber \\ 
&& +\sum_{i=2}^{m+1} \left( \begin{array}{c} m+1\\ i \end{array} \right)
   (-1)^{i+1}\,C(m+2-i,n-1) \nonumber \\
&& +(2n+m)\,C(m+1,n-1)+n\,(n-1)\,C(m,n-1) \Big] \nonumber \\ 
&& +B^+\,\sum_{i=1}^m \left( \begin{array}{c} m\\ i \end{array}
   \right)\,C(m-i,n+1)\;. \end{eqnarray} 
For a given spin $S$, this set of linear equations for the
correlations has to be solved for all $m+n\leq 2S+1$. The solutions
have to be put via equations (\ref{A5}) together with (\ref{A6}) into
equations (\ref{A33}), thus leading to a set of $2S$ equations for the
moments $\la(S^z)^p\ra$ ($p=1,\ldots,2S$), which have to be solved
self-consistently.  The highest moment $\la (S^z)^{2S+1}\ra$ has been
eliminated in favour of the lower ones through the relation
$\prod_{M_S}(S^z-M_S)=0$.

\end{document}